\documentclass[11pt]{article}

\usepackage{amsthm}
\usepackage{amssymb}
\usepackage{amsmath}
\newcommand{\myif}{\quad\text{if}\quad}

\DeclareMathAlphabet\EuFrak{U}{euf}{m}{n}       
\SetMathAlphabet\EuFrak{bold}{U}{euf}{b}{n}     

\begin{document}
\author{Sergio Doplicher
                         \\Dipartimento di Matematica
                         \\University of Rome ``La Sapienza''
                         \\00185 Roma, Italy  }

\title{The Measurement Process in Local Quantum Theory and the EPR Paradox}

\maketitle

\begin{abstract} We describe in a qualitative way a possible picture of 
the Measurement Process in Quantum Mechanics, which takes into account: 1. 
the finite and non zero time duration $T$ of the interaction between the 
observed system and the  {\itshape microscopic part}  of the measurement 
apparatus; 2. the finite space size $R$ of that apparatus; 3. the fact that 
the   {\itshape macroscopic part}  of the measurement apparatus, having 
the role of  {\itshape amplifying}  the effect of that interaction to a 
macroscopic scale, is composed by a very large but finite 
number $N$ of particles.

The conventional picture of the measurement, as an 
instantaneous action turning a pure state into a mixture, arises only in 
the limit $N \rightarrow  \infty, T \rightarrow 0, R \rightarrow  \infty$.

We sketch here a proposed scheme, which still ought to be made mathematically 
precise in order to analyse its implications and to test it  in specific models, where we argue that in Quantum Field Theory this picture should apply to the unique 
time evolution expressing the dynamics of a given theory, and should comply with the 
Principle of Locality. 

We  comment on the Einstein Podolski Rosen thought experiment \footnote{Partly modifying the discussion on this point in an earlier version of this note.}, reformulated here only in terms of  {\itshape local} observables (rather than {\itshape global} ones, as one particle or polarisation observables).The local picture of the measurement process helps to make it clear that there is no conflict with the Principle of Locality. 
\end{abstract}

\section{Introduction} 
The unitary, reversible nature of the Schroedinger time evolution of a 
quantum system seems foreign to the sudden irreversible jump from a  
{\itshape pure } state  $\omega$ (here a state is a positive linear normalised 
expectation functional over the algebra  $\mathfrak A$ of observables) to 
the  {\itshape statistical mixture}, where the expectation value of an observable $B$ is 
$\omega (E_1 B E_1 + ... + E_n B E_n) = \sum_j \omega (E_j) \omega_j (B)$ where $\omega_j (B) = \omega (E_j )^ {-1}     \omega (E_j B E_j) $, caused by the measurement of an 
observable (taken for simplicity with finite spectrum  
$\lambda_1,..,\lambda_n$) $A  = \sum_j  \lambda_j E_j $. The  {\itshape 
reduction of wave packets}  which occurs when one particular result  $\lambda_j$ 
is recorded, is supposed to take place instantaneously everywhere. We 
propose that this  {\itshape credo} should not be taken literally as an 
Axiom of Quantum Mechanics, but as an oversimplified limit of the picture 
of realistic interaction processes between the System $\mathcal S$ and the 
measurement apparatus $\mathcal A$. 

The first important step in the direction of reconciling the Measurement 
Process in Quantum Mechanics with the Schroedinger time evolution had been 
taken since the early years of Quantum Mechanics by John von Neumann 
\cite{dopl1}. The observed system $\mathcal S$ and the measurement apparatus $\mathcal A$ 
are described quantum mechanically by the tensor product of their Hilbert 
spaces of pure state vectors. The state vector of the composed system 
before the measurement is a product vector; the interaction between the 
two parts during the measurement takes that vector, by the unitary time 
evolution of the coupled system $\mathcal S$  +  $\mathcal A$, to another vector in the 
product Hilbert space, which is no 
longer a product vector, but a linear combination of such vectors. Therefore, the 
corresponding expectation functional (which is a pure state on the composed system) 
becomes a   {\itshape statistical mixture}  when restricted to the 
observables of the system alone.

Why have we to forget the observables of the measurement apparatus $\mathcal A$, 
thus loosing all the correlations between the possible outcomes? And how do 
we restrict to just one of those?
The answer von Neumann gave invoked the consciousness of the observer, a 
solution leading to well known paradoxes.

A scenario, which  in our view  is quite satisfactory, emerged first in 
a complete shape in 
the work of Daneri, Loinger and Prosperi \cite{dopl3} . The point, expressed in our own 
language, is that 
the measuring device $\mathcal A$ must consist of a {\itshape microscopic part}, that we will 
call $\mu \mathcal A$, and of a {\itshape macroscopic part}, that we will call $M \mathcal 
A$, which amplifies the results to our observed scale. The part $M\mathcal A$ interacts 
appreciably {\itshape only} with $\mu \mathcal A$, and is composed of $N$ 
particles.

As long as $N$ is kept finite, the whole system  $\mathcal S$ +  $\mu \mathcal A$ + 
$M \mathcal A$ has 
to be treated quantum mechanically, and the full state, if initially pure \footnote{This is definitely non-realistic if $N$ is very large.},  remains a pure 
state of the composed system under the Schroedinger evolution. But the
interference terms between the different outcomes will be so small, that the 
relative phases will be completely out of our experimental reach if $N$ is, 
as in practice, very large. Actually in concrete examples (bubble chamber, 
Geiger counters and their modern descendants, etc) the amplification is 
effectively an irreversible process (and as such   {\itshape in principle} 
reversible only if you take into account the fact that the number $N$ of 
the particles involved is finite; while we can well take it to be infinite 
for all practical purposes).

The wisdom on this point of view, built on previous arguments of Ludwig,
was later 
enriched considerably due to many works (by K.\ Hepp, D.\ H.\ Zeh, W.\ Zurek, 
G.\ C.\ Ghirardi, 
R.\ Gambini, and many others; we refrain from trying to give a complete 
reference list, which can be found e.g. in \cite{dopl4})  up to the 
recent work of G.\ Sewell \cite{dopl5}. Decoherence arises by the 
fast decrease with $N$ of the interference terms.

But while in the von Neumann picture one  {\itshape assumes} that there is an 
interaction potential between   $\mathcal S$  and  $\mathcal A$ which makes the 
measurement 
possible, in Quantum Field Theory the interaction is described from the 
start by the time evolution, as already discussed in some of the above 
references.

The only point of the present note is to suggest that, if we want to 
measure a {\itshape local observable} $A$, say localised in a space volume $V$ 
between time $t$ and $t + T$, then the operation of ``adding to a state the 
{\itshape microscopic part} $\mu \mathcal A$ of the apparatus which measures $A$'' 
should be {\itshape localised in the same }(or possibly in a larger but finite) 
{\itshape spacetime region};  while the amplifying {\itshape macroscopic part} 
$M\mathcal A$ will be localised in a much larger, macroscopic neighbourhood, which 
would be infinitely extended (at least in some direction) in the limit $N$ $ 
\rightarrow $ $ \infty $ (but practically extending e.g. from few microns to few 
centimetres, and from few microseconds to few seconds, around $V$ and $t$).
But the interaction between  $\mathcal S$ and 
$M\mathcal A$ 
will be neglected as inessential (the ionising cosmic ray crossing a 
bubble chamber does not interact with the bubble which develops as an 
effect of the ionisation of a single molecule, caused by one of its 
collisions), thus the decoherence, caused by the interaction between $\mu 
 \mathcal A$ and $M \mathcal A$, is  {\itshape not} the central point here.

At this point we take into account the {\itshape locality principle}, which 
requires that observables which can be measured in spacelike separated regions of 
spacetime should commute with each other; this is an expression of Einstein 
Causality: no physical effect can be transmitted faster than light. 

Accordingly, our picture of the measurement of an observable  $A$   {\itshape 
localised 
in some bounded spacetime region} entails that, if we test an arbitrary state with 
observables 
$B$ localised far away from that region  in a spacelike direction, we will find
{\itshape the same result} in the given state as in that 
obtained adding to that state the  {\itshape microscopic part}  $\mu \mathcal A$ 
(as well as adding also the  {\itshape macroscopic part} $M\mathcal A$ , if we are 
doing our test really far away) of the apparatus which measures $A$. Similar 
conclusions have been proposed long ago in the description of measurements in 
Local Quantum Field Theory by Hellwig and Kraus \cite{dopl6}.

In the case of the EPR gedanken experiment \cite{dopl2}, our discussion confirms that the entangled state, observed only in Alice's (resp.  only in Bob's) laboratory, is always the same mixed state, irrespectively of whether Bob (resp. or Alice)  performs his  (her) measurement in   
the spacelike separated laboratory.

But our picture makes it visible how the correlation between the two entangled possibilities, encoded in the preparation of the initial state, will manifest itself in the measurements, despite their spacelike separation. 

As the common wisdom of Local Quantum Theory suggests, there is no paradox: the existence of states with long range correlation is in no contradiction with locality, as shown explicitly by simple examples (Section 2).

These examples avoid the usual EPR picture involving one particle states and their polarisation: strictly speaking, a one-particle observable is  {\itshape never} a local observable, and  the angular momentum is anyway a global observable, as  the results of 
the Araki Yanase analysis \cite{dopl12} indicate.

We rather discuss the measurement of two non trivial projection operators, which are observables localised in two open bounded regions in spacetime, whose closures are spacelike separated.

In this case, an old result of Borchers shows that,  in particular, the intersection of the two projections, and that of their complements, are nonzero; and we  can choose unit vectors in the range of those intersections. It is then trivial to construct pure vector states of the quasilocal algebra which are superposition of two states, where the values of our projections are both zero in the first, both one in the second.

We emphasise that (at least for special choices of the projections, or, in reasonable theories, obeying the "Split Property", for any choice) the Hilbert space can be split in the tensor product of two factors, where each of our projections acts only on one of the two where the other one acts as the identity.  

In the tensor  product decomposition, our Hilbert space is exactly the (kinematical) description of two completely independent systems, and this state will exibit correlations, which might be very long range if our given spacelike separate regions are  very far away. Since the given theory is supposed to be an explicitly given   {\itshape local} theory (for instance, an explicit {\itshape free field} model of local algebras), no contradiction can emerge.

 \section{The pictures of the measurement 
process} 
Let $A  = \sum_j \lambda_{j} E_{j}$ be an observable of our system $\mathcal S$ 
with finite spectrum $  \lambda_{j} $, $j = 1,...,n$, and spectral 
selfadjoint projections $E_{j}$. The Hilbert space of the state vectors 
for the  composed system, consisting of the observed system $\mathcal S$ and the measurement 
apparatus $\mathcal A$, is the tensor product $\mathcal H_{\mathcal S} \otimes \mathcal H_ 
{\mathcal A}$ 
of 
the respective Hilbert spaces. 
In the von Neumann picture, the measurement of $A$ takes place from time 
$0$ to time $T$, due to the interaction $V$ between the system $\mathcal S$ and the 
measuring apparatus $\mathcal A$; the total Hamiltonian is 
\begin{equation}
\label{dopleq1}
 H =     H_ {\mathcal S}  \otimes I  + I \otimes  H_ {\mathcal A}   +  V;
\end{equation}
the duration $T$ is long enough for the measurement to take place, but so 
short (essentially instantaneous)  that we can neglect the evolution of  
$\mathcal S$ and  $\mathcal A$ on their own during the interval  from time $0$ to time 
$T$; 
there is a state vector $\Psi_  {R}$ describing the state ``ready to 
measure $A$'' of $\mathcal A$, such that the time evolution of the composed system   
from time $0$ to time $T$
has the effect
\begin{equation}
\label{dopleq2}
e^{iVT} E_{j} \Phi \otimes \Psi_  {R}   =     E_{j} \Phi \otimes \Psi_  {j},
 \end{equation}
where $\Phi$ is the initial state vector of  $\mathcal S$, and the $\Psi_  {j}$ are mutually orthogonal state vectors of $\mathcal A$. 
``Reading'' which of the  $\Psi_  {j}$  is the outcome,  we know the value of 
$A$ in the state of $\mathcal S$.

Thus on a generic state vector $\Phi$ of $\mathcal S$ the result of the 
interaction with $\mathcal A$ is
 \begin{equation}
\label{dopleq3}
e^{iVT}  \Phi \otimes \Psi_{R}   =    \sum_j E_{j} \Phi \otimes \Psi_{j}  
\equiv  \Psi'
 \end{equation}
so that, on any observable $B$ of $\mathcal S$, we have
\begin{align}
(\Psi', B \otimes I \Psi')  &=\nonumber  \sum_{j,k} ( E_{j} \Phi \otimes \Psi_  {j}, B 
\otimes I  E_{k} \Phi \otimes \Psi_  {k}) =\\
&=\nonumber \sum_{j,k} ( E_{j} \Phi , B E_{k} \Phi)  (\Psi_  {j}, \Psi_  {k})=\\ 
&= \sum_{j} (  E _  {j} \Phi , B  E _  {j} \Phi )\label{dopleq4}
\end{align}
which was the core of von Neumann's explanation why pure states are 
turned into mixtures.

 The operation of ``adding the apparatus to the state of our system'' is 
described by an isometry $W$ of  the Hilbert space $\mathcal H _{\mathcal S} $ of the 
system into  that of the composed system, $\mathcal H _{\mathcal S} \otimes \mathcal H _ 
{\mathcal A} $, given by
 \begin{equation}
\label{dopleq5}
W  \Phi  =   \Phi \otimes  \Psi_  {R} 
\end{equation}
which obviously commutes with all the observables of $\mathcal S$ \footnote{ More 
precisely, 
it 
is an 
intertwining operator between 
their actions on $\mathcal H _{\mathcal S} $ and on  $\mathcal H _{\mathcal S} \otimes 
\mathcal H_{\mathcal A} $.} ,  and which is changed by the 
evolution  from time $0$ to time $T$ to 
\begin{equation}
\label{dopleq6}
\alpha _{T} (W )       \equiv  e^{iHT}  W  e^{-iH_{\mathcal S} T} =  e^{iVT}  W =   
\sum_j W_  {j} E_  {j} ,
\end{equation}
where we turned to the Heisenberg picture, we neglected the independent 
evolutions of the system and of the apparatus over the time duration of the 
measurement, and we introduced the isometries
$W _{j} : \Phi \mapsto   \Phi \otimes   \Psi_  {j}$, which commute with all 
observables of $\mathcal S$   as well.

Now in the spirit of the Daneri, Loinger and Prosperi philosophy, we have 
to work rather with the Hilbert space  $\mathcal H _{\mathcal S} \otimes \mathcal H _ 
{\mu \mathcal A}   \otimes \mathcal H _{M\mathcal A} $ describing the composed system 
$\mathcal S$ + 
$\mu \mathcal A$ +  $M\mathcal A $, with the evolution governed by a Hamiltonian
\begin{equation}
\label{dopleq7}
 H =     H_ {\mathcal S}  \otimes I  \otimes I  +  I \otimes  H_ {\mu \mathcal A} \otimes 
I   +  V 
\otimes 
I  + I \otimes V'  + I \otimes I \otimes H_ {M \mathcal A}
 \end{equation}
where $V' $ describes the interaction of the macroscopic part of the 
measuring apparatus  {\itshape only} with the microscopic part (the interaction of $M \mathcal A$ with 
the system itself being neglected), interaction which is responsible for 
the amplification of the outcome $\Psi_  {j}$ to a state vector  $\Psi'_  
{j} (N)$ of $M\mathcal A $.

The crucial point for decoherence is that, in the limit $N \mapsto \infty$ 
of very large number of molecules composing the macroscopic part of the 
measuring apparatus, the vectors $\Psi'_  {j} (N)$  will tend weakly to 
$0$ \footnote{This supposes that the macroscopic part of the measuring apparatus with a given $N$  is a subsystem of those for any larger $N$.}, while the states they induce on the algebra of the macroscopic part 
of the measuring apparatus, will converge (in the weak* topology) to 
vector states of mutually disjoint representations.\footnote{That means representations 
of the algebra which cannot be connected by any nonzero intertwining 
operator. Note that we do not have to treat the macroscopic part of the 
measuring apparatus as a classical system in that limit: the parameters 
distinguishing different states in the thermodynamical limit (liquid 
versus gas in a bubble, etc) are classical parameters on their own, even if the 
description of the infinite system is quantum. } 

As those limiting states on the algebra of all observables of $M\mathcal A$ are 
disjoint from one another, tensoring them with vector states in the fixed 
representation on  $\mathcal H_{\mathcal S} \otimes \mathcal H_ {\mu \mathcal A} $ of the 
observables of the system plus the microscopic part of the measuring 
apparatus will give states $\phi_  {j}$ on the observable algebra of the 
three fold composed system which are pairwise disjoint as well. In any 
representation of that algebra, which contains among its vector states the 
states $\phi_  {j}$, linear combinations of the corresponding state 
vectors will thus induce a $mixed$ state, with no non vanishing term of 
interference between the components with different \(j's\) (for convenience of the reader, 
we reproduce the very elementary argument in the Appendix).

We want to memorise this essential role of $M\mathcal A$, but concentrate on the 
composed system $\mathcal S$ + $\mu \mathcal A$, on which the evolution describing the 
measurement takes pure states to pure states (more precisely, according to 
von Neumann, product state vectors to the entanglement of different 
product state vectors).

We now think of a Quantum Field Theory, which potentially describes all  
interactions which are relevant to the measurement of its observables, as 
already encoded in  {\itshape one and the same  } time 
evolution, and whose (vector) 
states include those describing the measurement apparatus (the microscopic 
part $\mu \mathcal A$, and, as long as $N$ is kept finite, $M\mathcal A$ as well; but in the 
limit where
$N$ goes to infinity, we would have to use vector states of disjoint 
representations). 

Thus  the Hilbert spaces  $\mathcal H_{\mathcal S} $,   $\mathcal H_{\mathcal S} \otimes 
\mathcal H_ {\mu \mathcal A}$, and 
  $\mathcal H_{\mathcal S} \otimes \mathcal H_ {\mu \mathcal A}   \otimes \mathcal H_ 
{M \mathcal A (N)} $ 
have to be identified with  {\itshape one and the same  }  Hilbert space  
$\mathcal H$, 
generated from the vacuum by the observables, or possibly that generated acting also with field 
operators which carry nonzero superselection charges.

 The operation of ``adding to a state the microscopic part of the measuring 
apparatus'' (as well as that pertaining to its macroscopic part, if $N$ is 
kept finite) are accordingly to be described by isometries of  
$\mathcal H$   {\itshape into itself }, say $W$ creating the state where 
``$\mu \mathcal A$ is ready to 
measure $A$'', and
 $W_  {j} $ creating the state ``$\mu \mathcal A$ has read the value $\lambda_  {j} $  
of $A$''.

With this proviso, their role should remain as in \eqref{dopleq6}, i.e., 
if $T$ denotes the duration of the measurement, $\alpha _  {t} (A) = U(t) A U(- 
t)$ the time evolution in the Heisenberg picture, in the spirit of the 
approximations above (the measurement is so fast that the evolution of the 
$E_  {j} $ in that time interval can be neglected), we have
\begin{equation}
\label{dopleq8}
 \alpha _{T} (W )    =   \sum_j W_  {j} E_  {j} ,\quad\quad         
W_{j}^{*}  W_  {k}  = \delta_{j,k} .
\end{equation}
In the  setting of (e.g. one particle) Quantum Mechanics discussed above, 
the $W$,  $W_{j}$ commuted with (better said, intertwined) all 
observables. If this were still the case in the present situation, they ought to 
change the vacuum state 
vector by a phase only. This is not possible, the result ought to be state
vectors describing $\mu \mathcal A$ alone, in the ``ready'' state or in the   state 
``$\mu \mathcal A$ has read the value $\lambda_  {j} $  of A''. Furthermore, $\mu 
\mathcal A$ 
(as well as $M\mathcal A$) has to be ``somewhere''.

Now we come to the crucial point. We are interested in a 
Local Quantum Field Theory  \cite{dopl7,dopl8} , where the observables form 
an algebra 
$\mathfrak A$  which is generated (technically as a C* Algebra) by the 
subalgebras   $\mathfrak A(\mathcal O)$, whose selfadjoint elements 
describe those observables which can be measured within the spacetime 
limits of the bounded region $\mathcal O$. 

The correspondence from bounded regions\footnote{This correspondence 
may be limited to nice regions as 
the intersections of a future and a past open light cone, which form a 
collection which is globally stable under Poincare' transformations, each 
region in the collection coinciding with the spacelike complement of its 
own  spacelike complement.} to algebras should preserve inclusions, span an 
irreducible algebra in the vacuum sector, be covariantly acted upon by the 
Poincare' (or at least the translation) group, fulfil the Spectrum Condition,
and, most 
notably, fulfil 
the   {\itshape locality postulate} , i.e. the 
measurements of two spacelike separated observables must be compatible, 
so that they commute with each other:
\begin{equation}
\label{dopleq9}
\mathfrak A(\mathcal O_1) \subset \mathfrak A(\mathcal O_2)' \myif 
\mathcal O_1 
\subset \mathcal O_2'                                      
\end{equation}
where the prime on a set of operators denotes its commutant (the set of 
all bounded operators commuting with all the operators in the given set) 
and on a set in Minkowski space denotes the spacelike complement.

At least in theories where there are no massless particles, we can without 
any loss of generality suppose that  $\mathfrak A$ is included in a larger 
algebra  $\mathfrak F$  of fields, with a similar structure of local 
subalgebras, each generated by field operators fulfilling normal 
Fermi/Bose commutation relations at spacelike separations, according to 
whether the superselection sectors they reach acting on the vacuum 
are both paraFermi or not \cite{dopl9}  (the notion of  {\itshape Statistics of a 
Superselection Sector} being intrinsically defined solely in terms of the local 
observables \cite{dopl10,dopl11}; for an expository account, see e.g. 
\cite{dopl19}, \cite{dopl20}). 

These fields  
can be assumed to generate, acting on the vacuum state vector,  {\itshape all} the 
superselection sectors of the theory (defined as the collection of all representations 
of  $\mathfrak A$ which describe, in an appropriate precise sense, elementary 
perturbations of the vacuum). All this is just encoded in the $observable$ algebra 
$\mathfrak A$.

Now, if we think of the measurement process of a $local$ observable, say 
$A$ in  $\mathfrak A(\mathcal O)$, (for simplicity taken with finite 
spectrum as above), we expect that the measurement process is {\itshape local }! 

It is 
then natural to assume that the isometries $W ,    W_  {j}$  fulfilling  
\eqref{dopleq8}, {\itshape are elements of the same local subalgebra }  
$\mathfrak A(\mathcal O)$ of observables, or   of fields, $\mathfrak 
F(\mathcal O)$. At the only price, of course, of replacing  $\mathcal O$ 
by a slightly larger region which contains, together with  $\mathcal O$, 
its time translates by amounts not exceeding T.

We might well call ``locally measurable observables'' those local 
observables for which this is possible. (But it might also happen that the 
$W , W_ {j}$ have to be localised in a larger region; or they might fail 
to be exactly isometric; or they might be quasilocal and not strictly 
local\footnote{The important point is that they should create, acting 
on the vacuum, a state essentially localised in the desired region 
$\mathcal O$. An 
isometry from $\mathfrak
F(\mathcal O)$ does that exactly; an operator from $\mathfrak
F(\mathcal O)$, normalised so that its action on the vacuum gives a unit 
vector, does that with the better approximation the closer to one is 
its norm; without the last condition, it might give a result as close 
as we want to any vector, by the Reeh - Schlieder Theorem  \cite{dopl7}.}. 
In 
the last events, most conclusions would hold only asymptotically, by the 
cluster property).

For example if we think of the measurement of a (suitable spectral projection of 
a) charge density smeared over a small spacetime neighbourhood, mimicked, in 
practice, say by a single ionisable molecule in the metastable liquid in a bubble 
chamber, then $W = W_{0}$ and $W_{1}$ would be the creation operators from the 
vacuum of the strictly localised states of the appropriate baryon number, 
describing in a reasonable approximation the non ionised respectively ionised 
molecule, localised in $\mathcal O$.~\footnote {The field algebras 
$\mathfrak F (\mathcal O)$ are actually {\itshape generated } by isometric field 
operators which, acting on the vacuum, create strictly localised states with 
specific values of the superselection quantum numbers  \cite{dopl9}. But, strictly 
speaking, such an exactly 
localised state ought to have a completely indefinite number of particles, and, in 
order to describe with a reasonable approximation the state of a single molecule, 
it ought to be localised in a region not smaller than some minimal size; moreover, 
for this example to be correct, we ought to leave out from the description of the 
time evolution precisely the electromagnetic forces; otherwise, the localisation 
could only be approximate; replacing the relevant identities with approximate ones 
which become more and more exact after finite but larger and larger space 
distances, however, should not 
alter the essence of our point.}

Of course, the $W$'s cannot commute with all observables. In analogy with 
the non local picture of non relativistic Quantum Mechanics, they should 
lie in a subalgebra of  $\mathfrak A(\mathcal O)$ (or of   $\mathfrak F (\mathcal O)$; more precisely,  they should lie in a type $I_\infty$ 
subfactor) commuting with the measured observable $A$. 

Similarly, if we want to take care of the amplifying part of the 
measurement apparatus, we ought to apply to the state vector where we want 
to measure $A$, not only $W$, but also some operator $W'(N)$ (localised in some 
larger macroscopic region including $\mathcal O$~\footnote{\ldots and extending to 
infinity, at least in some direction, when the number $N$ of atomic 
constituents of the amplification device tend to infinity.}). But for all 
practical purposes a macroscopic region of the size few microns to few 
centimetres will suffice: that is to say, it is practically pointwise if we think 
of 
astrophysical separations.

In quite the same way as mentioned above, the interaction between the  $ W_  {j}$'s and the  $W'(N)$'s will 
explain the decoherence, occurring exponentially for large $N$'s.

But  the $W$'s are localised in $\mathcal O$; if we add to any state our 
apparatus $\mu \mathcal A$, the state vector, $\Phi$, is changed  to $W \Phi$; 
the expectation value of any observable $B$ which is localised in a 
spacelike separated region {\itshape will not change}:
\begin{equation}
\label{dopleq10}
 (W\Phi, B  W \Phi)  =  (\Phi, W^{*} B  W \Phi) =    (\Phi, B W^{*}  W \Phi) =    
(\Phi, B  \Phi),
\end{equation}
where we used the local commutativity of $W$ and $B$, and the isometric 
nature of $ W $, i.e. $ W^{*} W = I$.

Since, by our choice of a slightly larger $\mathcal O$, also $\alpha _{T} (W 
)$ is localised in   $\mathcal O$, the same applies after the 
measurement took place.

We conclude that the state described by  the vector  $ \Phi $ , whether or 
not we add to it the microscopic measurement apparatus evolved  in time within some interval which includes that of the 
measurement of $A$, {\itshape looks exactly the same to an observer which is 
spacelike away} from our region $\mathcal O$. If that observer were 
$macroscopically$ far away in a spacelike direction, the same would be 
true even if we added also the amplifying part of the measurement apparatus - 
provided we kept the number of its micro-constituents finite.

However, this does not imply that the effects of long range entanglement, possibly present in a given state, have to disappear. We discuss explicitly this point in a typical example. 

Consider two 
non trivial (neither $0$ nor 
$I$) selfadjoint projections (``questions'')  $E$ and $F$ which are 
localised in regions which are {\itshape well}   spacelike separated from each 
other \footnote{This means that the modulus of the (negative) Minkowski 
distance between their points has a  non zero lower bound.}. By a result 
of H.J.Borchers \cite{dopl14}, $EF$  
and $(I - E) (I - F)$ can't be $0$; if  $\Psi$ and  $\Phi$ are unit vectors in the 
range of  $EF$  and $(I - E) (I - F)$ respectively,  the unit vector  $\Psi^{*}   =  2^{-1/2}(\Psi  +  \Phi)$  induces, on the whole 
quasilocal 
algebra, a pure state where the values of   $E$ and $F$ are entangled. Of course, 
no contradiction with local commutativity can arise from the existence of such a state.

Let $W_{E}, W_{F}$ denote, as above,  the  creation operators of  ``the microscopic part of the measuring apparatus'' which performs  the measurement of $E$ in the first region, resp. of $F$ in the second. 

The measurement of both $E$ and $F$ in the state described by the unit vector  $\Psi^{*}$ will be described by the action of the isometry $W_{E} W_{F}$ on the vector  $\Psi^{*}$ (insertion of the microscopic part of the two measurement apparatuses) and, after the measurements took place, by 
\begin{align}
\nonumber
 \alpha _{T} (W_{E} W_{F})\Psi^{*}& = \sum_j W_  {j, E} E_  {j}  \sum_j W_  {j', F} F_  {j'}\Psi^{*} =\\ 
&=  \sum_j W_  {j, E} E_  {j}  \sum_j W_  {j', F} F_  {j'}  2^{-1/2}(\Psi  +  \Phi)
\label{dopleq12}
 \end{align}
  where $j, j'$ run from $1$ to $2$, and $E_1 = E, E_2 = I - E$, and similarly for $F$.

 Now, since our two measurements are supposed to take place in mutually spacelike separated regions, the $  E_  {j}  $ and the $ W_  {j', F} $ commute with one another, and all the mixed terms in the double sum vanish since, by the choice of $ \Psi^{*}$, 
 \begin{equation}
\label{dopleq13}
 (I - E) F \Psi^{*}  = E(I -  F) \Psi^{*} = 0
 \end{equation}
 and the state vector above takes the form
 \begin{equation}
\label{dopleq14}
2^{-1/2} (W_  {1, E} E_  {1}  W_  {j1, F} F_  {1}\Psi  +   W_  {2, E} E_  {2}  W_  {j2, F} F_  {2} \Phi)
\end{equation}

This elementary fact is crucial: these relations make explicit the existence of the entanglement of our states both prior or after the insertion of the two ({\itshape microscopic}) measurement apparatuses. Nevertheless, due to Locality, each of the two measurements has no consequence which is detectable in the other region.

Moreover, if the two regions were macroscopically apart, and we meant to have included in the creation operators $W_{E},  W_{F}$  also the description of the macroscopic part of the measurement apparatus, the result would not have changed: again, by local commutativity, observations performed in only one of the two regions,  could detect no change in the state.

The same would apply if we performed the limit where the number of microscopic constituents of one or both the amplifying devices tends to infinity.

The entanglement will be revealed only by joint measurements in the two regions, but would never be visible when only the restriction of our state to the algebra of observables localised in only one of the two regions is considered: this restriction will be one and the same   {\itshape mixed} state irrespectively whether the measurement in the spacelike separated region is performed or not \footnote{Note that, due to the structure of the von Neumann algebras  $\mathfrak A(\mathcal O)$, {\itshape no} vector state can be a pure state when restricted to  $\mathfrak A(\mathcal O)$. But the point here is the bilocalised nature of our state. Otherwise, as it appears in the subsequent discussion, in interesting theories it would be enough to consider a type $I$ factor $N$ containing   $\mathfrak A(\mathcal O)$ and contained in the local algebra of a just slightly larger region, and plenty of vector states would have a pure state restriction to $N$ - but never to   $\mathfrak A(\mathcal O).$ }.

 For the same reason, of course, decoherence will occur if  {\itshape at least one}  of the two observables is measured, inserting the microscopic {\itshape and} macroscopic parts of the measurement devices (exactly in the same way as, in our discussion earlier on in this Section, decoherence of the final states of $M\mathcal A$ causes decoherence of the final states of  $\mathcal S$.  

We close this section with a description of more special choices of our example of states exhibiting correlations of local observables. In these examples, there will be  a decomposition of the given Hilbert space as the tensor product of three factors, where the third is merely a spectator, and the first two describe, with their tensor product, the composed states of two independent systems. The observables of these subsystems will be special collections of observables localised in one of the two regions.

The situation is then exactly the same as in the von Neumann picture of the measurement process: pure (vector) states of the composed system will typically induce mixed states of a single system. 

Thus the mathematical picture of the EPR thought experiment is exactly the same as for two generic independent systems; and the correlations initially prepared in our given state will remain, without any contradiction with locality. 

This version of the EPR thought experiment has the advantage of avoiding the contradiction between the local character of the observed quantities and the fact that  observables referring to  {\itshape a single photon} (or a single particle in general) {\itshape cannot be local}.

Moreover, while  vector (or "normal") states restricted to a local algebra  $\mathfrak A(\mathcal O)$ 
cannot be pure nor can they have any discrete decomposition into pure states, we will describe explicitly the possibility to detect  {\itshape locally} the correlations in our model state. 

This possibility is seen more clearly in a theory fulfilling the "Split Property" (see  \cite{dopl20} and References in there). In such a theory, whenever $\mathcal O _ {a}$ and $ \mathcal O _ {b}$ are nice regions (open double cones) the second containing the closure of the first, then there exists a type $I$ factor $N$ such that
 \begin{equation}
\mathfrak A(\mathcal O_a ) \subset  N  \subset \mathfrak A(\mathcal O_b ).
 \end{equation}
Thus, if  $\mathcal O_1 ,  \mathcal O_2 $ are {\itshape well} spacelike separated double cones, there will be type $I$ factors $N_1 , N_2$ such that 
\begin{equation}
\mathfrak A(\mathcal O_j ) \subset  N_j ,   j  =  1,2, 
 \end{equation}
 with  $N_1 , N_2$ still contained in the subalgebras of two mutually spacelike separated double cones, and hence commuting. They will generate a  type $I$ factor $M$ consisting of localised operators, in a pair of regions slightly larger than $\mathcal O_1,  \mathcal O_2 $, and contained in the algebra of some large double cone containing $\mathcal O_1$ and $ \mathcal O_2 $.

 The Hilbert space will then decompose into a tensor product  $ \mathcal H_1   \otimes  \mathcal H_2   \otimes  \mathcal K$, so that  $N_1 , N_2$ and $M$ identify respectively with all bounded operators on $ \mathcal H_1 ,   \mathcal H_2 , \mathcal H_1   \otimes  \mathcal H_2 $, tensored with the identity operators on the remaining factor, i.e.  $ B (\mathcal H_1 )  \otimes  I \otimes I$, $ I \otimes B( \mathcal H_2 ) \otimes I$ and $B( \mathcal H_1   \otimes  \mathcal H_2 ) \otimes I $ respectively.

  Let now  $\Psi_  {j}$ and  $\Phi_  {j}$, $j = 1,2$ be unit vectors 
respectively in the Hilbert spaces  
$ \mathcal H_1 $ and in   $\mathcal H_2 $, in the range of $E$ for $j = 1$ and of $I - E$ for $j = 2$, resp. of $F$ for $j = 1$ and of $I - F$ for $j = 2$, and consider the pure state on all bounded 
operators of the tensor product  $ \mathcal H_1   \otimes  \mathcal H_2    \otimes \mathcal K$induced by the vector
 $(1/2^ {1/2}) ( \Psi_{1}\otimes \Phi_{1} + \Psi_{2}\otimes\Phi_  {2}) \otimes \Phi$, where $\Phi$ is any unit vector in $\mathcal K$. 

On the quasilocal algebra it will define a state where the values of the local projections in $\mathcal O_1 $,  $\mathcal O_2 $ we chose are correlated, whatever the distance between $\mathcal O_1 $,  $\mathcal O_2 $ is. The mathematical expression of that state is exactly the same as for the example referring to two independent systems.

This state will describe correlation which are not detectable by separate observations in only one of the regions  $\mathcal O_1,  \mathcal O_2 $; but its restriction to $M$ will be the pure state $\eta$ of  
 $B( \mathcal H_1   \otimes  \mathcal H_2 )$ induced by the unit vector  $(1/2^ {1/2}) ( \Psi_{1}\otimes \Phi_{1} + \Psi_{2}\otimes\Phi_  {2})$. Therefore the correlations will be accessible to appropriate local measurements which are localised in our two regions.

The typical (and most accurate) examples of such an observable are obtained as follows.

 Let $Q$ denote the orthogonal projection in $ \mathcal H_1   \otimes  \mathcal H_2 $ onto the one dimensional subspace spanned by $(1/2^ {1/2}) ( \Psi_{1}\otimes \Phi_{1} + \Psi_{2}\otimes\Phi_  {2})$; then, in the above tensor product decomposition, $Q \otimes I$ belongs to $M$, hence is an element (a selfadjoint idempotent) $B$ of a  local algebra of observables localised in the union of two regions slightly larger than  $\mathcal O_1,  \mathcal O_2 $ respectively.

 Obviously measuring $B$ one detects the correlations built into our state.

More specifically, by measuring $B$ in any state $\omega$ produces the mixture assigning to the observable $A$ the value $\omega (BAB  +  (I - B) A (I - B)) $, and keeping the $\omega (B)$ fraction of cases where "$B$ has the value $1$" one prepares a state $\phi$ such that, for any $A$, we have $\phi (A)  =  \phi (BAB)$. Since for each $C$ in $M$ we have $BCB  =  \eta (C) I$ (i.e. a number times the identity operator), this means that $\phi  =  \eta$ when restricted to $M$, and a fortiori  {\itshape when restricted to the algebra of observables localised in our two regions.} This possibility of  {\itshape local preparation of states} is actually equivalent to the "Split Property"  \cite{dopl16}. 
 
 Thus, while a separate measurement in  $\mathcal O_1$  or  $ \mathcal O_2 $, will never reveal  
the correlations present in our state, a suitable localised measurement will do. 

Another possibility is to check the violation of Bell's inequality, as achieved in remarkable experiments  \cite{dopl13}.
  
  Eventually, we point out that a similar discussion is possible without necessarily requiring the Split Property. For, the algebras $\mathfrak A(\mathcal O)$ of observables localised in the bounded region $\mathcal O$ will contain lots of type  $I_\infty$ subfactors  \footnote {Indeed, in physically meaningful theories they will be of type$III_1$ \cite{dopl21}.}. 

 To see this in a very elementary way, choose a non unitary isometry $W$ in   $\mathfrak A(\mathcal O)$ (by the quoted result 
of H.J.Borchers \cite{dopl14}, every nontrivial projection localised in a region well contained in $ \mathcal O $ will be the range of such an isometry). Then the Wald decomposition of $W$ into the direct sum of a unitary operator and the multiple of a shift will provide a projection $G$, the strong limit  of $W^{n}  {W^{n}}^*$ (hence lying in $\mathfrak A(\mathcal O)$), such that the restriction of $W$ to the range of $ I - G$ is a multiple of a shift; hence the weakly closed *subalgebra $M$ generated by $W(I - G)$ is the direct sum of a type  $I_\infty$  factor on  the range of $ I - G$ and zero on the orthogonal complement.

If we chose $W$ from the start localised in a region well contained in $\mathcal O$, then $I - G$ would be the range of an isometry $S$ in $\mathfrak A(\mathcal O)$, and  $S^* M S$, the dilation of the corner of M to the whole Hilbert space, is a  type  $I_\infty$  factor.  

\section{Concluding remarks} 

The conventional picture of the measurement process in Quantum Mechanics, 
as an instantaneous jump from a pure state to a mixture, which affects the 
state all over space at a fixed time in a preferred Lorentz frame, 
appears, in the scenario we outlined, as  the result of several limits:

1. the time duration $T$ of the interaction giving rise to the measurement 
(which, in an exact mathematical    
treatment, would  involve the whole interval from  minus infinity  to plus infinity, as  all 
scattering processes) is set 
 equal to zero;

2. the number of microconstituents of the amplifying part of the measurement apparatus is set equal to infinity, thus allowing {\itshape exact } decoherence;

3. the volume involved by the measurement apparatus in its interaction 
with the system (thus  occupied by the microscopic part of the apparatus) tends to the whole 
space, allowing the reduction of wave packets to take place {\itshape everywhere}.

One should compare the necessity of the last limit with the results of 
the Araki Yanase analysis \cite{dopl12} , whereby the additivity of angular momentum is 
compatible with the conventional description of the measurement process 
only in the limit where the measurement involves a volume  extended to the 
whole space.

Strictly speaking, taking those limits independently from one another does not seem physically 
possible: if we are interested in the limit where decoherence becomes exact, 
the time needed for the evolution of  $M\mathcal A$  to one of the final 
disjoint states has to be taken into account as well, and that time, though in practice negligeably small, strictly speaking will be 
infinite (especially in a relativistic theory, since if  $M\mathcal A$ 
contains infinitely many microscopic constituents, it has to be infinitely 
extended, and the return to equilibrium would require infinite time).

The discussion above is but a qualitative picture of a possible scenario;  to 
make it a bit more precise, many mathematical questions should be answered: one should replace
\eqref{dopleq8} by a suitable approximation,  and investigate the nature of the convergence 
for large time values, in order to justify the assumption that we can pick a 
finite 
value of $T$ as a reasonable approximation. Furthermore, the mathematical meaning 
of the three limits 
above should be studied carefully.

But,  in the above picture, the interaction described by the time 
evolution is  {\itshape the same} as the interaction between our system and the 
apparatus; one could expect that, e.g., in a free field theory there 
should be, in the sense described here, no observables at all! 

More generally, in a full and precise theory of the measurement process in Local Quantum Physics, the subset of those observables 
localised in a region  $\mathcal O$  which are ``locally measurable''
in the same region  $\mathcal O$ might be a proper subset; it would 
coincide with {\itshape all} the observables localised in that same region only in
model theories which offer a description of a {\itshape complete set of 
interactions}. 

It would be extremely interesting to characterise such situation as a kind of  {\itshape ergodicity property  of the 
interactions}. But in order to try and formulate such a property, first one would need a precise mathematical definition of the creation operators, denoted $W$ and $W'$ in the previous discussion,  of the (approximate) local measurement devices,  for the microscopic part and the
amplifying macroscopic part respectively, whose role was described here purely in intuitive terms.
 
 \section{ Appendix.  Disjoint states cannot be superposed  }
We reproduce, for the convenience of the reader, the very elementary argument which prevents 
superposition of two states $\phi$ and $\psi$ if they induce disjoint representations of a C* 
algebra $\mathfrak A$.

Let $\pi$ be a representation of $\mathfrak A$ on a Hilbert space $ \mathcal H $, 
and $\xi$, $\eta$ two vectors in $ \mathcal H $ which induce $\phi$ and $\psi$ as 
vector states of the representation $\pi$. For instance, $\pi$ could be the direct 
sum of the two GNS representations. Let $E$, $F$ denote the orthogonal projections 
on the closed subspaces obtained applying the whole representation to the vectors 
$\xi$, $\eta$ respectively. These subspaces are stable under the representation 
hence these projections commute with it.  The product $EF$ is a linear bounded 
operator from the range of $F$ to the range of $E$, and commutes with the 
representation $\pi$, since $E$ and $F$ do. Their ranges are invariant subspaces 
for the representation $\pi$, hence
 $EF$ is an intertwining operator between the restrictions of $\pi$ to those ranges.
 
 But those restrictions are unitarily equivalent to the GNS representations of $\phi$ and 
$\psi$, which we assumed to be disjoint. Thus those restrictions are disjoint, and they have 
 no nonvanishing intertwining operator. We conclude that $EF = 0 $.

Now let $\zeta = a \xi + b \eta $ be a (normalised) linear combination of our 
vectors; it induces a vector state $\rho$ on $\pi$ whose value on any element $A$ 
of $\mathfrak A$ is 
\begin{align}\nonumber
\rho&(A)  = (\zeta, \pi (A) \zeta)  = ( a \xi + b \eta , \pi (A) a \xi + b \eta )\\\nonumber
&= a^{*}a( \xi , \pi (A)  \xi ) + b^{*}b ( \eta , \pi (A)  \eta )  
+ a^{*}b ( \xi , \pi (A)  \eta )  + ab^{*} ( \eta , \pi (A)  \xi )=\\\nonumber 
&= a^{*}a( \xi , \pi (A)  \xi ) + b^{*}b ( \eta , \pi (A)  \eta )\\ 
&= (|a|^{2} \phi + |b|^{2} \psi)(A), 
\label{dopleq11}
\end{align}
where we used the fact that the last two terms in the expansion of the scalar 
product, the {\itshape interference terms},  vanish for all $A$ in $\mathfrak A$, 
since $E \xi = \xi, F \eta = \eta$ by  construction, $E$ and $F$ commute with 
$ \pi (A) $, and $EF = 0 $. Therefore, 
for any choice of the representation vectors of our states, the attempt to 
construct a superposition fails, and gives only a statistical mixture.

{\itshape Acknowledgements}. It is a pleasure to thank Gherardo Piacitelli for 
discussions and valuable comments. We are also grateful to Chris Fewster for drawing our attention to some related references.

\end{document}